\documentclass[preprint2]{aastex}

\newcommand{\be}{\begin{equation}}
\newcommand{\ee}{\end{equation}}

\newcommand{\ax}{$\alpha_{\rm X}$}

\newcommand{\rb}[1]{\raisebox{1.5ex}[-1.5ex]{#1}}
\newcommand{\msun}{$M_{\odot}$}

\newcommand{\plm}{$\pm$}

\newcommand{\swift}{{\it Swift}}
\newcommand{\xmm}{{\it XMM-Newton}}

\shorttitle{Swift monitoring of Mkn 335}
\shortauthors{Grupe et al.}

\begin{document}



%
%
%


\title{  A remarkable long-term light curve, and deep, low-state  
spectroscopy:
  \swift\ \& \xmm\ monitoring of the NLS1 galaxy Mkn 335}


\author{Dirk Grupe\altaffilmark{1},
S. Komossa\altaffilmark{2,3,4},
Luigi C. Gallo\altaffilmark{5},
Anna Lia Longinotti\altaffilmark{6,7}
Andrew C. Fabian\altaffilmark{8},
Anil K. Pradhan\altaffilmark{9},
Michael Gruberbauer\altaffilmark{5},
Dawei Xu\altaffilmark{10}
}

\altaffiltext{1}{Department of Astronomy and Astrophysics,  
Pennsylvania State
University, 525 Davey Lab, University Park, PA 16802;
grupe@astro.psu.edu}

\altaffiltext{2}{Technische Universit\"at M\"unchen, Fakult\"at f\"ur  
Physik, James-Franck-Strasse 1/I, D-85748 Garching, Germany; stefanie.komossa@gmx.de}

\altaffiltext{3}{Excellence Cluster Universe, TUM, Boltzmannstrasse 2,  
85748 Garching, Germany}

\altaffiltext{4}{Max Planck Institut f\"ur Plasmaphysik,  
Boltzmannstrasse 2, 85748 Garching, Germany}

\altaffiltext{5}{Department of Astronomy and Physics, Saint Mary's  
University, Halifax, NS
B3H 3C3, Canada; lgallo@ap.stmarys.ca}

\altaffiltext{6}{MIT Kavli Institute for Astrophysics and Space Research
77 Massachusetts Avenue, NE80-6011 Cambridge, MA 02139}

\altaffiltext{7}{XMM-Newton Science Operations Centre ESAC,
P.O. Box 78 28691 Villanueva de la Ca\~nada, Madrid, Spain}

\altaffiltext{8}{Institute of Astronomy, Madingley Road, Cambridge,  
CB3 0HA, UK}

\altaffiltext{9}{Department of Astronomy, The Ohio State University,  
140 W 18th Av, Columbus, OH; pradhan@astronomy.ohio-state.edu}

\altaffiltext{10}{National Astronomical Observatories, Chinese 
Academy of Sciences, 20A Datun Road, Beijing
100012, China; dwxu@nao.cas.cn}




\begin{abstract}
The Narrow-line Seyfert 1 galaxy (NLS1)
Mkn 335 is remarkable because it has
repeatedly shown deep, long X-ray low-states which show pronounced
spectral structure. 
It has become one of the prototype AGN in deep
minimum X-ray states.
Here we report on the continuation of our ongoing monitoring campaign with
\swift\ and the examination of the low state X-ray spectra based on a 200 ks
triggered observation with \xmm\ in June 2009.
\swift\ has continuously monitored Mkn 335 since May 2007 typically on a
monthly basis. This is one of the longest simultaneous UV/X-ray light curves
so far obtained for an active galactic nucleus (AGN). 
Mkn 335 has shown strong X-ray 
variability even on time scales of hours. In the UV,
it turns out to be one of the most variable among NLS1s.
Long-term Swift monitoring allow us to examine correlations between the UV,
 X-rays and X-ray hardness ratios.  We find no significant correlation or 
 lag between the UV and X-ray variability; however, we do find distinct 
 trends in the behavior of the hardness ratio variability.  The hardness ratio and count rate are correlated in the low-flux state, but no correlation is seen in the high-state.
The X-ray low-state spectra of
the 2007 and 2009 \xmm\ observations display significant spectral
variability.
We fit the X-ray spectra with a suite of phenomenological models in order to characterize the data.
The broad band CCD spectrum can be fitted equally well with partial absorption and blurred reflection models.  These more complicated models are explored in further detail in upcoming work.

\end{abstract}


\section{Introduction}

While AGN typically vary in X-rays by  factors of a 
few \citep[e.g.][]{grupe01},
some AGN show dramatic drops in their X-ray flux, accompanied by changes in their spectrum.
These  AGN are known to be typically X-ray bright,
  but for some time they display very low flux states which makes them  
different
  from AGN such as PHL 1811 which are intrinsically X-ray weak  
\citep{leighly07}.
Some recent examples of such deep minimum states observed in
AGN are PG 2112+059, PG 1535+547, PG 1543+489, RX J2217.9--5941, Mkn  
335,
PHL 1092, and PG 0844+349
\citep{schartel10, schartel07,
ballo08, vignali08, grupe04, grupe07b, grupe08a, miniutti09, gallo10}.
Absorption has always been considered to play an important role
in explaining AGN X-ray spectra. 
Variability through a variable absorber may play a much more
important role in AGN
than previously thought \citep[e.g.][]{turner09}. Some of the best  
examples of
variable absorbers are e.g. NGC 1365 \citep{risaliti09},  Mkn 766 \citep{miller07,
turner07}, 1 H0557--385 \citep{longinotti09}, and NGC 3516 \citet{turner11}.

However, this picture is far from being complete and clear. 
Besides absorption, a popular
explanation of the X-ray low states in AGN are reflection models such as  
proposed by
\citet{fabian02, fabian04, fabian09} for 1H 0707-495, or by  
\citet{gallo06} for NLS1s in general. Both, reflection and partial  
covering
absorber models produce relatively similar X-ray spectra \citep[see e.g.  
discussion in
][]{grupe08a}. To make things even more complex, as shown by e.g.  
\citet{chevalier06} and
\citet{merloni06}, the X-ray spectrum can be dominated in the high state by  
reflection
and is then modified by a partial covering absorber when the AGN is in  
a low
state. Even for MCG-6-30-15, which has been the poster-child for  
reflection
models after the ASCA detection of a broad red wing of the Fe K$\alpha
$ line by
\citet{tanaka95}, \citet{miller08, miller09} argued that its X-ray
spectra can be consistently explained by partial covering absorption.

Bright AGN in deep low-states with well-covered light curves and 
good low-state spectra are essential for further exploring the physics
which are responsible for the structures and features seen in AGN
X-ray spectra, and especially in low-states, where spectral complexity
is most pronounced. 
The NLS1 Mkn 335
($\alpha_{2000}$ = $00^{\rm h} 06^{\rm m} 19.^{\rm s}5$,
$\delta_{2000}$ = $+20^{\circ} 12' 11 \farcs 0$) is such
an AGN: it goes repeatedly
into deep low-states, is relatively X-ray bright even in 
those states, and has been monitored by us for years to identify these low-states.   
It shows interesting spectral structures, and one possible interpretation
has been that it exhibits an unusually broad Fe line \citep{longinotti07a}. 
It is nearby ($z$=0.0258), well studied in the optical spectral band, 
and has a well-measured
BH mass of $1.4 \times 10^{7} M_{\odot}$ from reverberation 
mapping \citep{peterson04, grier11}.

Mkn 335 has been  
known to be an X-ray bright
AGN for decades, starting with UHURU \citep{tananbaum78} and EINSTEIN  
\citep{halpern82}.
EXOSAT and BBXRT observations suggested a strong soft X-ray excess in  
the X-ray spectrum of Mkn
335 \citep[][respectively]{pounds87, turner93} while \citet{nandra94}  
reported on the
presence of a warm absorber in this source. During the ROSAT All Sky-
Survey and pointed
observations Mkn 335 appeared to be X-ray bright, and was modeled with a strong soft X-
ray excess
\citep{grupe01}. From the 1993 observations by ASCA \citep{george00},  
\citet{lei99b} concluded
that the X-ray spectrum was affected by the presence of a warm  
absorber, while
\citet{ballantyne01} interpreted the spectral shape by X-ray  
reflection on the accretion
disk. Mkn 335 was also observed by \xmm\ in 2000 and 2006
\citep{gondoin02,longinotti07a,longinotti07b,oneill07,arevalo08} and {\it  
Suzaku} in June 2006
\citep{larsson08}. In all cases Mkn 335 was X-ray bright,
with the exception of one EXOSAT observation in 1983 \citep{pounds87}.

However, when
Mkn 335 was observed by \swift\ \citep{gehrels04}  in 2007 May
as part of a \swift\ fill in project to study the spectral energy  
distributions
in AGN \citep{grupe10} it appeared to
be dramatically fainter in X-rays than expected from  
previous X-ray observations \citep{grupe07b}. In order to investigate the
nature of the low-state in more detail we initiated a Target-of-
Opportunity (ToO)
observation with \xmm, which was executed on 2007 July 10  
\citep{grupe08a}. During
this 22 ks \xmm\ observation we discovered strong soft X-ray 
emission lines of  H and He-like ions such as OVII \citep{grupe08a, longinotti08} in
the Reflection Grating Spectrometer  \citep[RGS; ][]{denherder01}.  
These lines are
only visible during an extreme X-ray low-state and they can provide  
information of the
physical conditions of the gas surrounding the central black hole.  
Because the 22ks
\xmm\ observation was too short to obtain any reliable line ratios, we  
applied for a 200ks
\xmm\  observation which would be triggered by a  
low state seen by
\swift. When \swift\ started the monitoring campaign again in May 2009  
after Mkn 335
came out of the sun constraint, it appeared to be again in an extreme low  
state. We therefore
triggered our approved \xmm\ observation and observed Mkn 335 for 200 ks in 2009   
June 11th to
14th.   Mkn 335 also became a
target of our \swift\ Guest Investigator program in 2008 in which we  
monitored the
AGN on a weekly basis in X-rays and all 6 UVOT filters.

Here we report on the results of the \swift\ monitoring campaign of  
Mkn 335 and the
continuum short term light curve measured by \xmm\ during the 200 ks triggered
observation.  
This first paper in a sequence focusses on presenting the rich data sets, 
on simple modeling, and on revealing spectral trends.  
In-depth modeling of the multi-component warm absorber based on
the RGS data,  and a detailed investigation of blurred reflection models
 will each be presented in follow-up work. 
This Paper is organized as follows: in Section 2 we describe the  
observations and
the data reduction of the \swift\ and \xmm\ observations. In Section 3  
we
present the long- and short-term light curves and the analysis of the  
X-ray
spectra obtained by \xmm. The results of this analysis are  
discussed in Section 4.
Throughout the paper spectral indices are denoted as energy spectral  
indices with
$F_{\nu} \propto \nu^{-\alpha}$. Luminosities are calculated assuming  
a $\Lambda$CDM
cosmology with $\Omega_{\rm M}$=0.27, $\Omega_{\Lambda}$=0.73 and a Hubble
constant of $H_0$=75 km s$^{-1}$ Mpc$^{-1}$ corresponding to  a  
luminosity distance D=105 Mpc.
All errors are 90\% confidence unless stated otherwise.

\section{\label{observe} Observations and data reduction}

\subsection{Swift observations}

\swift\ started monitoring Mkn 335  on 2007 May 17 and still  
continues with
this campaign on at least a monthly basis (Table\,\ref{swift_log}), except for  
the period of 
February to May when Mkn 335 is in sun constraint for \swift.
As part of a Swift Guest
Investigator program the cadence was changed to once per week starting  
in June 2008.
  The X-Ray Telescope \citep[XRT, ][]{burrows04}
  observations were performed in Photon Counting mode  \citep[PC mode]
[]{hill04}.
X-ray data were reduced with the task {\it xrtpipeline} version 0.12.1.
Source and background photons were extracted
with {\it XSELECT} version 2.4, from
circles with radii of 47$^{''}$ and 189$^{''}$, respectively when the  
source count rate was less than
0.4 counts s$^{-1}$. However, during some parts of our monitoring  
campaign the count rates were significantly
higher than 0.4 counts s$^{-1}$ which means that the data were  
affected by pileup. In order to avoid the effects of
pileup we excluded the inner part of the Point Spread Function,  
depending on the brightness of the AGN.
  The spectral data were re-binned with at least 20 photons per bin
using {\it grppha} version 3.0.0.
The 0.3-10.0 keV spectra were
analyzed with {\it XSPEC} version 12.3.1x \citep{arnaud96}.
The auxiliary response files were created with {\it
xrtmkarf} and corrected using the exposure maps,
and the standard response matrices {\it  
swxpc0to12s0\_20010101v011.rmf} and
{\it swxpc0to12s6\_20010101v011.rmf} were used for the observations  
before and after
the XRT substrate voltage change in August 2007, respectively  
\citep{godet09}.

The UV-optical Telescope \citep[UVOT,][]{roming04} covers the range  
between
1700-6500\AA\ and  is a sister instrument  of \xmm's OM. Although it has
a similar set of filters to the OM \citep{mason01, roming04},
the UVOT UV throughput is a factor of about 10 higher than that of the
OM.
The UVOT data were coadded for each segment in each filter with the UVOT
task {\it uvotimsum} version 1.3.
Source photons in all filters
  were selected in a circle with a radius of 5$^{''}$.
  UVOT magnitudes and fluxes were measured with the task {\it  
uvotsource} version
  3 based on the most recent UVOT calibration as described in  
\citet{poole08} and
  \citet{breeveld10}.
The UVOT data were corrected for Galactic reddening
\citep[$E_{\rm B-V}=0.035$; ][]{sfd98}. The correction factor in each  
filter was
calculated with equation (2) in \citet{roming09}
who used the standard reddening correction curves by \citet{cardelli89}.

\subsection{XMM-Newton observations}

We observed Mkn 335 with \xmm\ \citep{jansen01}
on 2009 June 11 to 14 for a total of 200 ks split over orbits 1741 and 1742.
A summary of the
observations with each of the instruments on-board \xmm\ is given in
Table\,\ref{xmm_log}. The European Photon Imaging Camera (EPIC) pn
\citep{strueder01} was operated in Large Window  mode with the thin
filter. This combination was chosen to avoid pileup in case the AGN
re-brightened. The two EPIC MOS \citep{turner01} were both operated in  
Full-Frame
mode with the medium filters.
High-resolution X-ray spectroscopy was performed
using the two Reflection Grating Spectrometers \citep[RGS; ][]
{denherder01}
on-board \xmm.  Optical photometry was performed in 5 filters with
the Optical Monitor \citep[OM; ][]{mason01}.  The data are used
to measure the optical-to-X-ray spectral energy distribution of Mkn
335 during the \xmm\ observation. Due to slew problems at the  
beginning of the
observations, V-filter observations were not obtained. All OM  
observations were
performed in a science-user defined configuration with a 7$^{'}\times  
7^{'}$
observing window.

The \xmm\  data were processed in the standard way using the
XMMSAS version {\it xmmsas\_20100423\_1803-10.0.0}.
The EPIC pn data were checked for episodes of high particle background.
At the end of the first orbit (odsID 0600540601) the pn data were  
strongly affected by
high particle background. Times with a background at energies E$>$ 10  
keV was larger
than 10 counts s$^{-1}$ were
screened and not used for spectral analysis.
  This left an effective observing time of 99036s. During the second  
orbit
  (ObsID 0600540501) there were only very short episodes of high  
particle background.
  The total screened exposure time during this orbit was 69339s.

The source X-ray photons in the EPIC pn and MOS
were selected in a circular region with a radius of 1$^{'}$.  Likewise,
background photons were selected from a nearby, source-free region
with the same radius. Only single and double events ({\tt PATTERN.le.4})
and single to quadruple events ({\tt PATTERN.le.12})
were selected
  for the pn and MOS data, respectively.
The spectra were rebinned with the XMMSAS task {\it specgroup} with an  
oversampling of 3
of the  resolution elements at the energy of the bin.
The redistribution matrices and the auxiliary response files were  
created by the
XMMSAS tasks {\it rmfgen} and {\it arfgen}, respectively.
We included also the 2007 \xmm\ pn data in our analysis. Note,  
however, that we also applied
{\it specgroup} to rebin this spectrum and that the results may  
slightly differ from those
presented in \citet{grupe08a}. For comparison purposes, we also
display the 2006 \xmm\ high-state data in form of a light curve and
hardness ratios.

RGS spectra and response matrices were created by the standard RGS  
XMMSAS tool
{\it rgsproc}. The RGS spectra were rebinned with 10 photons per bin  
using
{\it grppha}. Spectral fits to the EPIC pn and MOS, and RGS spectra  
were performed with XSPEC
version 12.3.1x \citep{arnaud96}.
The OM data were processed with the XMMSAS task {\it omichain}. The  
magnitudes and fluxes
of Mkn 335 were taken from the source lists created by the {\it  
omichain} task.
For the count rate to flux conversion we used the conversion factors  
given in
the OM Calibration document {\tt XMM-SOC-CAL-TN-0019}.

\subsection{Xinglong optical spectroscopy  \label{xinglong}}

The optical spectrum of Mkn 335 displays strong high-ionization 
iron coronal lines. In order to search for changes in the broad emission
lines, and in the coronal lines,  we have triggered an optical spectroscopic
observation of Mkn 335 with the 2.16m Xinglong telescope quasi-simultaneous
with the 2009 \xmm\ observation.     

The data were acquired on 2009 July 31 with the Opto-Mechanics Research
(OMR) spectrograph
equipped with a 600 line mm$^{-1}$ grating and the 2$''$ slit. 
This setup produces a resolution of 5\AA. 
The spectrum of Mrk 335 was taken with 3600s exposure. 

Data reduction was done following standard procedures 
using IRAF. The CCD reductions included bias subtraction, 
flat-field correction, and cosmic-ray removal.
Wavelength and flux calibration were performed.

We find that within the uncertainties there is 
 no variability in the coronal lines, similar to
our previous result \citep{grupe08a}.

\section{\label{results} Results}

\subsection{Long-term light curve observed by \swift \label{var_long}}

\begin{figure*}
\epsscale{1.5}
\plotone{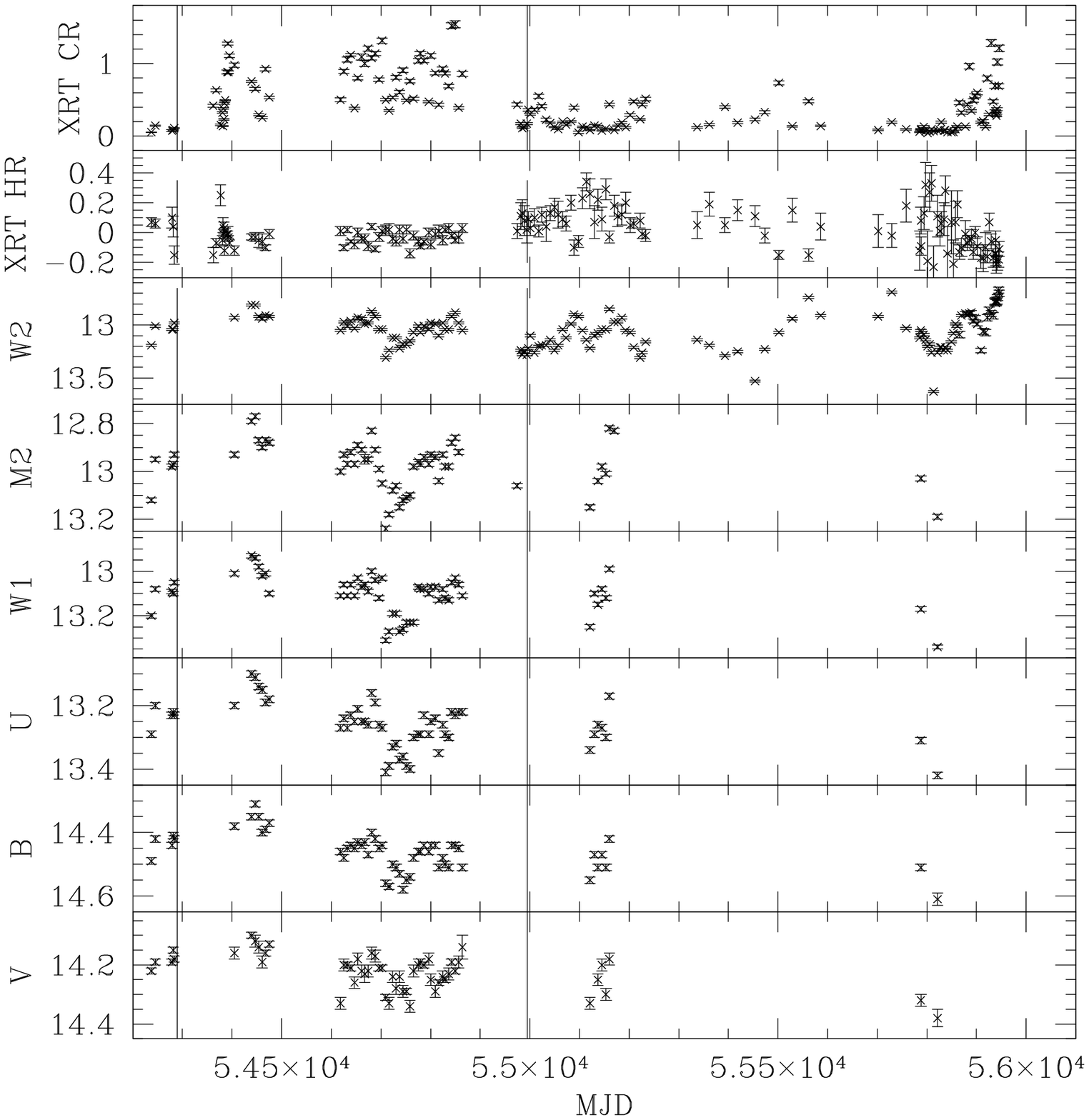}
\caption{\label{mkn335_lc} \swift\ XRT and UVOT light curves of Mkn 335. The vertical 
lines at MJD 54290 and 54995 mark the times of the \xmm\
observations in July 2007 and June 2009.  The beginning of the light curve is
2007 May 17.
}
\end{figure*}

\begin{figure*}
\epsscale{2.6}
\plottwo{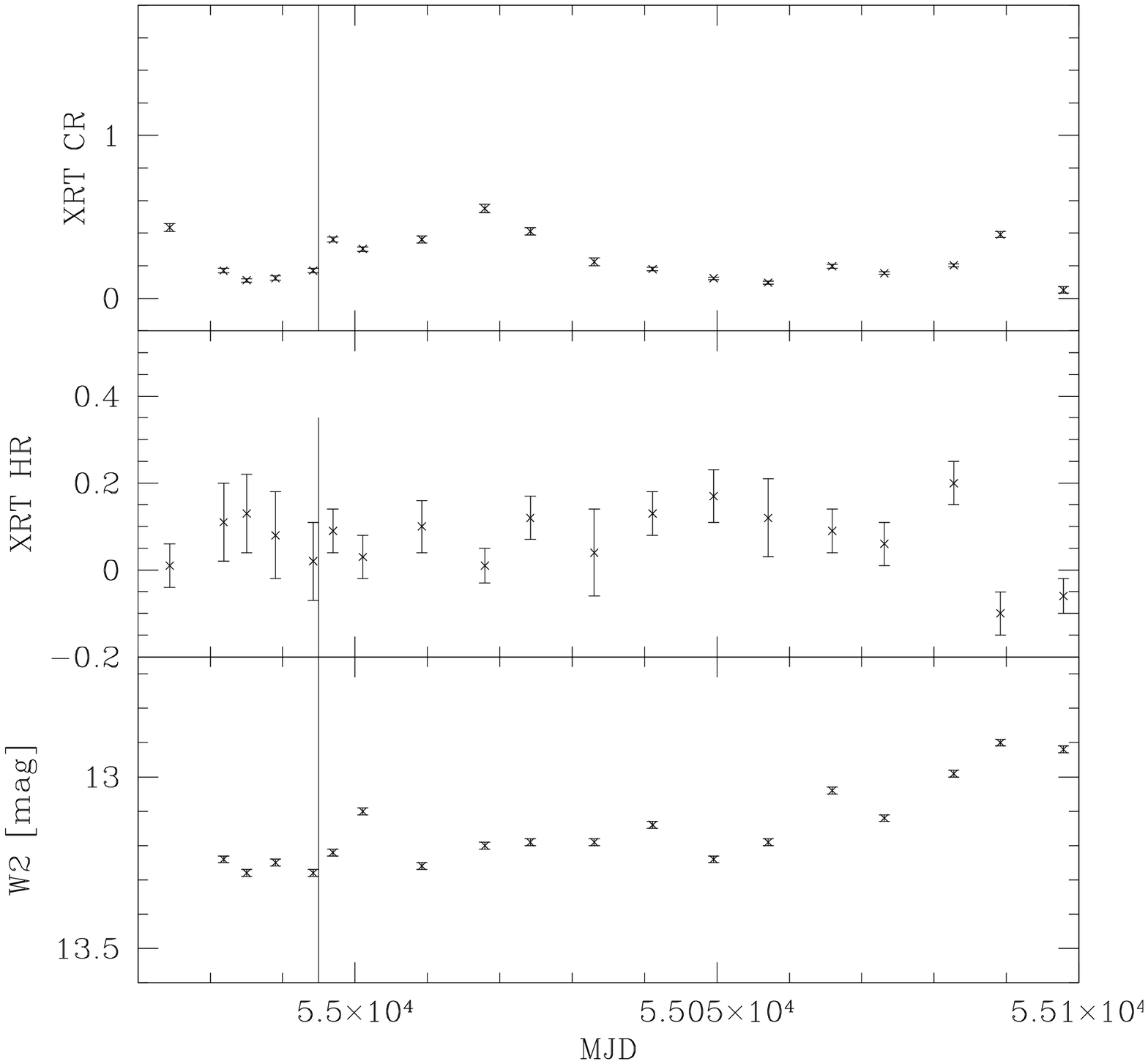}{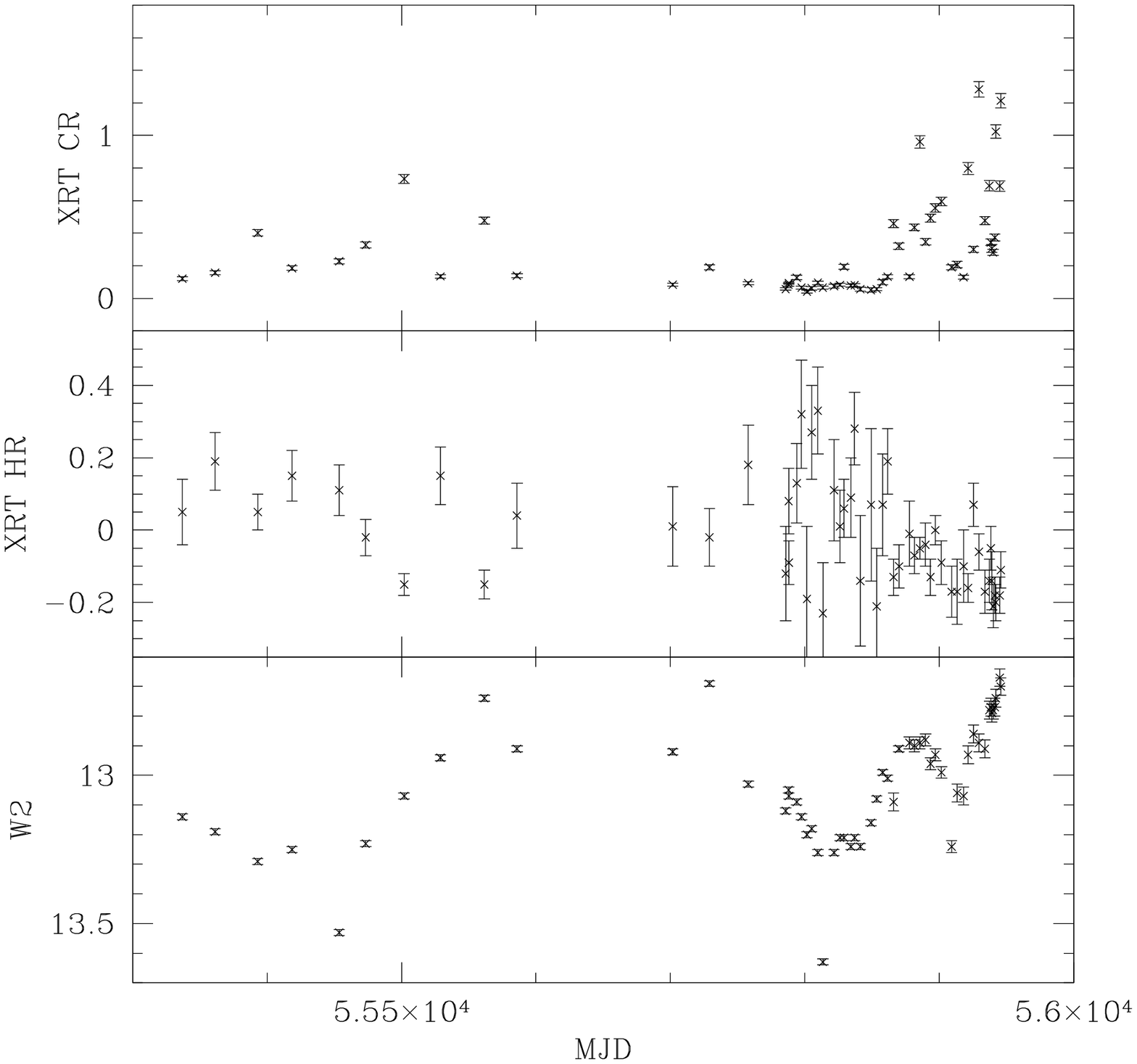}
\caption{\label{mkn335_lc_2009} Zoom-in of the  \swift\ XRT and UVOT W2
light curves of Mkn 335.
The left panel displays the \swift\ observations before and shortly after the 2009 \xmm\ observation. 
The vertical line at MJD 54995 marks the time of the \xmm\
observation in  June 2009.  The beginning of this light curve is 2009 May 23. The right panel shows the light curves in
 2010 and 2011
}
\end{figure*}

Figure\,\ref{mkn335_lc} displays the \swift\ XRT count rate and  
hardness ratio light
curves as well the the UVOT light curves in each of the 6 filters.  
Note that after the
end of the \swift\ GI program for Mkn 335 in January 2009, we limited  
the UVOT observations
to W2  in order to reduce the UVOT filter wheel rotations. The  
vertical lines in Figure\,\ref{mkn335_lc}
mark the times of the 2007 and 2009 \xmm\ observations.
  The XRT count rates, hardness ratios and UVOT magnitudes for these  
light
curves are summarized in Table\,\ref{swift_results}. Note that the  
2009 \xmm\ observation
was performed from MJD 54993.3188 to 54996.2500. Compared with the  
time at which we
triggered the 2009 \xmm\ observation, Mkn 335 had become significantly  
brighter and 
 we found Mkn 335 in an interesting transition into an intermediate flux state.
It increased its average 
XRT count rate from about 0.11 to 0.36 counts s$^{-1}$ during the time of the \xmm\ observation.
The left panel in 
Figure\,\ref{mkn335_lc_2009} displays the \swift\ XRT and UVOT W2  
light curves
before and after the 2009 \xmm\ observation.

\begin{figure}
\epsscale{1.0}
\plotone{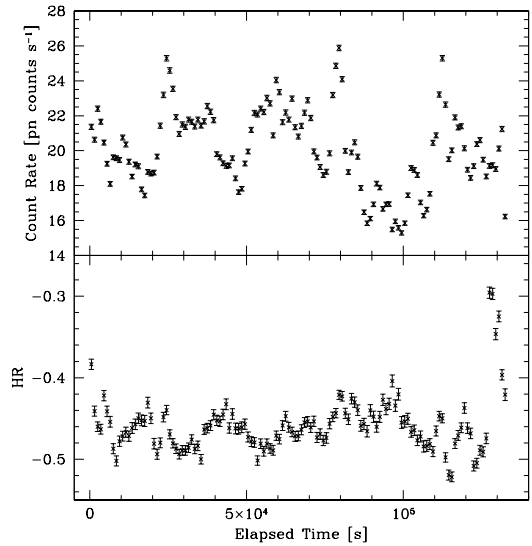}
\caption{\label{mkn335_xmm_lc_2006} \xmm\ pn  light curves of Mkn 335 during the
2006 observation.
}
\end{figure}

When we started monitoring Mkn 335 in May 2007 it was in its  
historical low state as we
reported in \citet{grupe07b}. However, it became significantly  
brighter starting from
September 2007 and Mkn 335 remained in this intermediate state  
throughout 2008
(Figure\,\ref{mkn335_lc}).
Due to a failure of one of the \swift\ gyros \citep{grupe07b} the UVOT  
was turned off in September and
October 2007. When we started monitoring Mkn 335 again after it came  
out of the
\swift\ sun-constraint in May 2009,
  we found Mkn 335 back in a low state. This low state was the reason  
why we
triggered our pre-approved \xmm-observation. Mkn 335 remained in a low state
throughout 2009 with
a slight increase towards the end of that monitoring episode.
In 2010 Mkn 335 has been in a low state for
most of the time. After emerging from the \swift\ sun-constraint in May 2011,  it
shows 
again a very low state with a \swift\ XRT count rate of 0.08 counts
s$^{-1}$  and even displayed an all-time low state on 2011 August 28 with a
count rate of 0.042\plm0.010 counts s$^{-1}$ (right panel in Figure\,\ref{mkn335_lc_2009}).
However, recently in November 2011 Mkn 335 went into a high state peaking at about 1 count s$^{-1}$
in the XRT.
 We therefore changed our observing strategy to a four day cadence. Mkn 335 is now showing 
a very rapid variability behavior suggesting that it switches into a high state. Currently (January 2012) we
are observing Mkn 335 daily. On 2012 January 11 it displayed the highest XRT count rate measured since January 
2009 with 1.3 counts s$^{-1}$. 
This X-ray flux is comparable to the 2006 \xmm\ observation shown 
in Figure\,\ref{mkn335_xmm_lc_2006}.

\begin{figure}
\epsscale{0.7}
\plotone{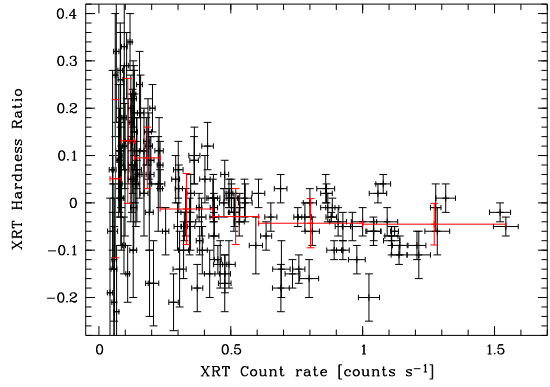}
\caption{Count rate vs. hardness ratio during 
 the \swift\ observations. The red
crosses correspond to bins in count rate containing 25 measurements and the mean
hardness ratio and standard deviation in that bin.
\label{mkn335_xrt_cr_hr}
}
\end{figure}

One question regarding a highly variable source like Mkn 335 is, does the  
spectral shape change with X-ray flux?
As we have shown already in \citet{grupe08a} the X-ray spectra of Mkn  
335 look completely different in the low and high states.
The low number of photons in the \swift\ XRT spectra, however,
does not allow us to perform a detailed spectral analysis. Still, a hardness ratio  
provides some
clues about the changes in the X-ray spectrum. Figure\,
\ref{mkn335_xrt_cr_hr} shows the relation between the count rate and
hardness ratio in the \swift\ data\footnote{We define the hardness  
ratio as HR = (H-S)/(H+S) with S and H being the background corrected
counts in the 0.3-1.0 and 1.0-10 keV bands, respectively}.
Figure\,\ref{mkn335_xrt_cr_hr} suggests that the AGN becomes softer  
with increasing count rate.
This results is a linear correlation coefficient $r_{\rm l}=-0.580$  
and a
Spearman rank order correlation coefficient
$r_s=-0.69$ with a Student's T-value $T_s=-9.4$. For both correlations  
the probability of a random results is P$<10^{-6}$.

As shown in Figure\,\ref{mkn335_lc},
Mkn 335 also shows variability in all 6 UVOT filters. 
During the 2009 \xmm\ observation, Mkn 335 appeared to be slightly  
fainter by about 0.2-0.3 mag in
the OM B, U, W1, M2, and W2 Filters (Table\,\ref{xmm_log}) compared  
with the 2007 observation. This is in agreement with the \swift\
UVOT light curves  displayed in Figure\,\ref{mkn335_lc}.
As listed in Table\,\ref{swift_results}, in the UVOT W2 filter Mkn 335  
was about 0.3
magnitudes brighter during the May 2007 X-ray low state compared with  
the 2009 May/June low-state. UV variability by 0.3 mag is quite common among NLS1s
\citep{grupe10}. However, Mkn 335 exhibited a remarkable change in W2 in 2010 
as displayed in the right panel of Figure\,\ref{mkn335_lc_2009}
when 
after a sudden drop from 13.25  
to 13.54 in September 2010 it became brighter by
0.79 mag over a period of just three and a half months.
As shown in the right panel of Figure\,\ref{mkn335_lc_2009}, it shows an even stronger drop between June and September 2011. 
On 2011 June 11 it reached its brightest UV state  seen during our 
entire \swift\ monitoring campaign with UVW2=12.69 mag. 
 This brightening,
however, was followed by a continuous fading in the W2 filter with the 
lowest measurement with
13.63 mag on 2011 September 09. This is a drop in W2 by almost 1 magnitude 
within three months,  
equivalent to an increase in flux
by a factor of  2.5 - one of the strongest  
changes in the UV observed in our entire
AGN sample \citep{grupe10}, which even exceeds the UV variability seen in 
WPVS 007 \citep{grupe07a, grupe08b}. 
These drops in the UV flux are extremely rare in NLS1s, which tend to show 
no or only little UV variability \citep[e.g.][]{grupe10}.  
Now in January 2012 Mkn 335 displays a very high flux in the UV again, peaking on 
January 18 with a magnitude of 12.67 in UV W2.

\begin{figure}
\epsscale{0.7}
\plotone{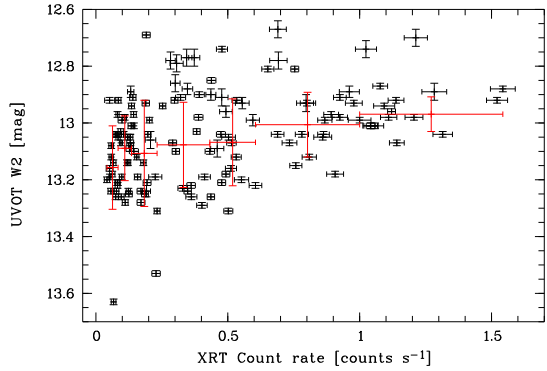}
\caption{\swift\ XRT Count rate vs.  UVOT W2 magnitude. The red
crosses correspond to bins in count rate containing 25 
measurements and the mean
UVOT W2 magnitude and standard deviation in that bin.
\label{mkn335_swift_cr_w2}
}
\end{figure}

  Figure\,\ref{mkn335_swift_cr_w2} displays the relation between the XRT
count rate and the UVOT W2 magnitude. This plot shows  that Mkn 335 is  
only found to be faint in the UV
when the AGN is faint in X-rays. However, it appears to be bright in  
the UV independently of the XRT count rate.
We found a linear correlation coefficient $r_{\rm l}=-0.376$ with a
probability P=0.00020 of a random result.
A Spearman rank order test results in a correlation
coefficient $r_s=-0.347$, $T_s=3.50$ with a probability P=0.00072 of a
random result. However, there is not a direct correlation with the  
source being
bright in the UV, when also being bright in X-rays. There is a large scatter  
in the UV W2 magnitude when the
AGN appears to be X-ray faint. Note that the faintest and the  
brightest UV W2 magnitudes are both measured when
Mkn 335 has a count rate less than 0.5 count s$^{-1}$. The large  
scatter may suggest that the X-ray and UV emission 
generally do not vary together.

To explore this in more detail, 
we investigated potential lags in the variability
of the \swift\ XRT count rate  and the UVOT
W2 filter data during the time period 2007 May tp 2011 July.
Using the Bayesian framework \citep[e.g.][for an  
introduction to
Bayesian analysis]{gregory05, albert09}
we created ``synthetic" model count rates from the observed
W2 magnitudes. We first transformed
the magnitudes into relative fluxes and then applied a Gaussian
bandpass filter in order to only retain variability on specific time
scales of interest.
For every time scale that was tested the same filter was also applied
to the observed count rates. The filtered ``synthetic" time series was
free to be rescaled and shifted in flux, as well as translated in
time.
Overall, our model has 4 free parameters: time lag, flux offset and
linear scaling factor, and width of the Gaussian used for the
timescale of the bandpass filter. We assigned uniform prior
probabilities within sensible parameter ranges for all parameters.
Assuming normally distributed uncertainties, we then calculated the
likelihood of obtaining the observed count rate data, given a particular set
of parameter values for our model. Using the nested sampling code
MultiNest \citep{feroz09},
we calculated the posterior probability
distributions for all 4 parameters, as well as the Bayesian evidence.
For comparison, we also calculated the Bayesian evidence for a
reference model with constant flux (i.e., no information from the W2
magnitudes is used except for the temporal sampling). Our calculations
show that the constant flux model is preferred over the model created
from the W2 magnitudes.
Therefore, we conclude that at present there is not enough evidence
for lags between the XRT count rate and W2 filter for Mkn 355. A more detailed
description of the method used, and its application to the Mkn 355
data, will be
provided in an upcoming paper (Gruberbauer et al., in preparation).

\subsection{Short-term variability observed by \xmm \label{var_short}}

\begin{figure}
\epsscale{1.0}
\plotone{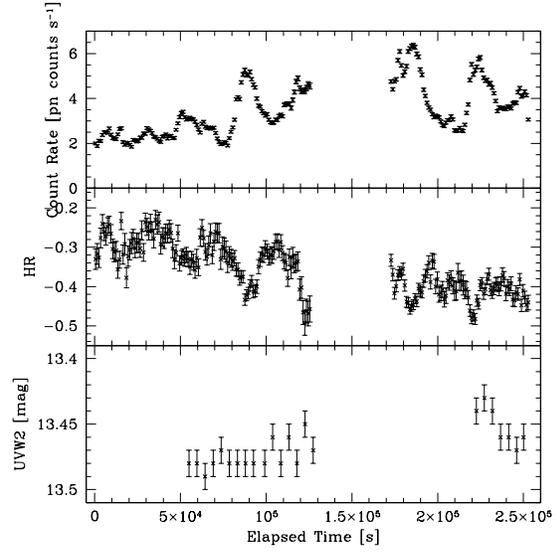}
\caption{\label{mkn335_xmm_lc_2009} \xmm\ pn and OM W2 light curves of Mkn 335 during the
2009 observation.
}
\end{figure}

Figure\,\ref{mkn335_xmm_lc_2009} shows the \xmm\ EPIC pn count rate  
and hardness ratio and
OM W2 light curves. The pn light curve was binned in 1000s bins.
The W2 bins are typically 4400 s as listed in
Table\,\ref{xmm_log}. Overall,
Mkn 335 appeared to be brighter during the 2009 June observation  
compared with the
2007 July observation \citep{grupe07b}.
The overall trend is that the AGN becomes softer when the overall  
count rate increases
from the beginning to the end of the \xmm\ observations. Also note  
that the `flares`
  appear to be soft. These 'flares' show doubling times of roughly 3  
hours.
This `flaring' is similar to what had been reported by  
\citet{oneill07} during
the 133 ks 2006 \xmm\ high-state observation.
The light curve from the 2006 \xmm\ observation is displayed in
Figure\,\ref{mkn335_xmm_lc_2006} for comparison.

\begin{figure*}
\epsscale{1.5}
\plottwo{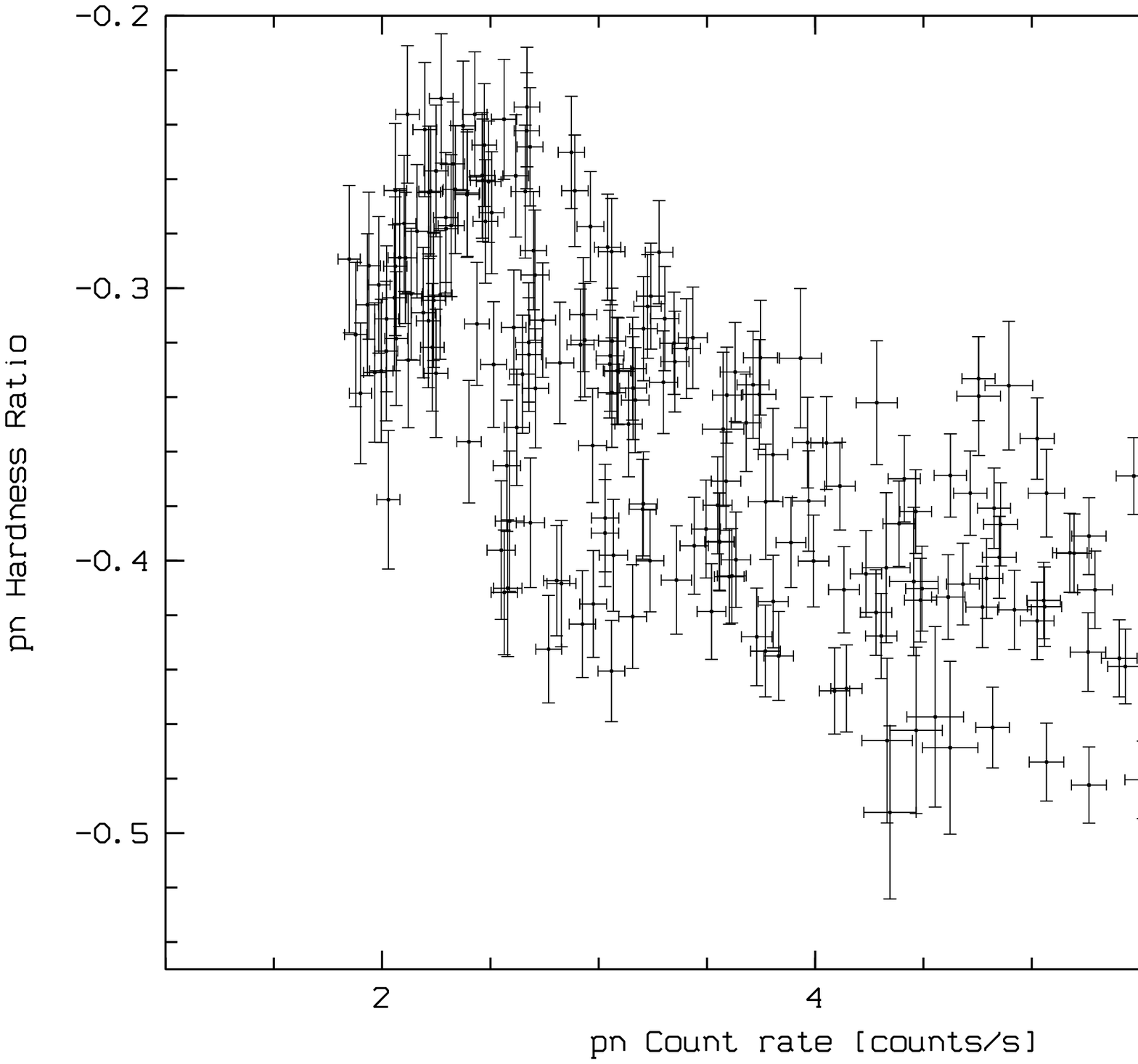}{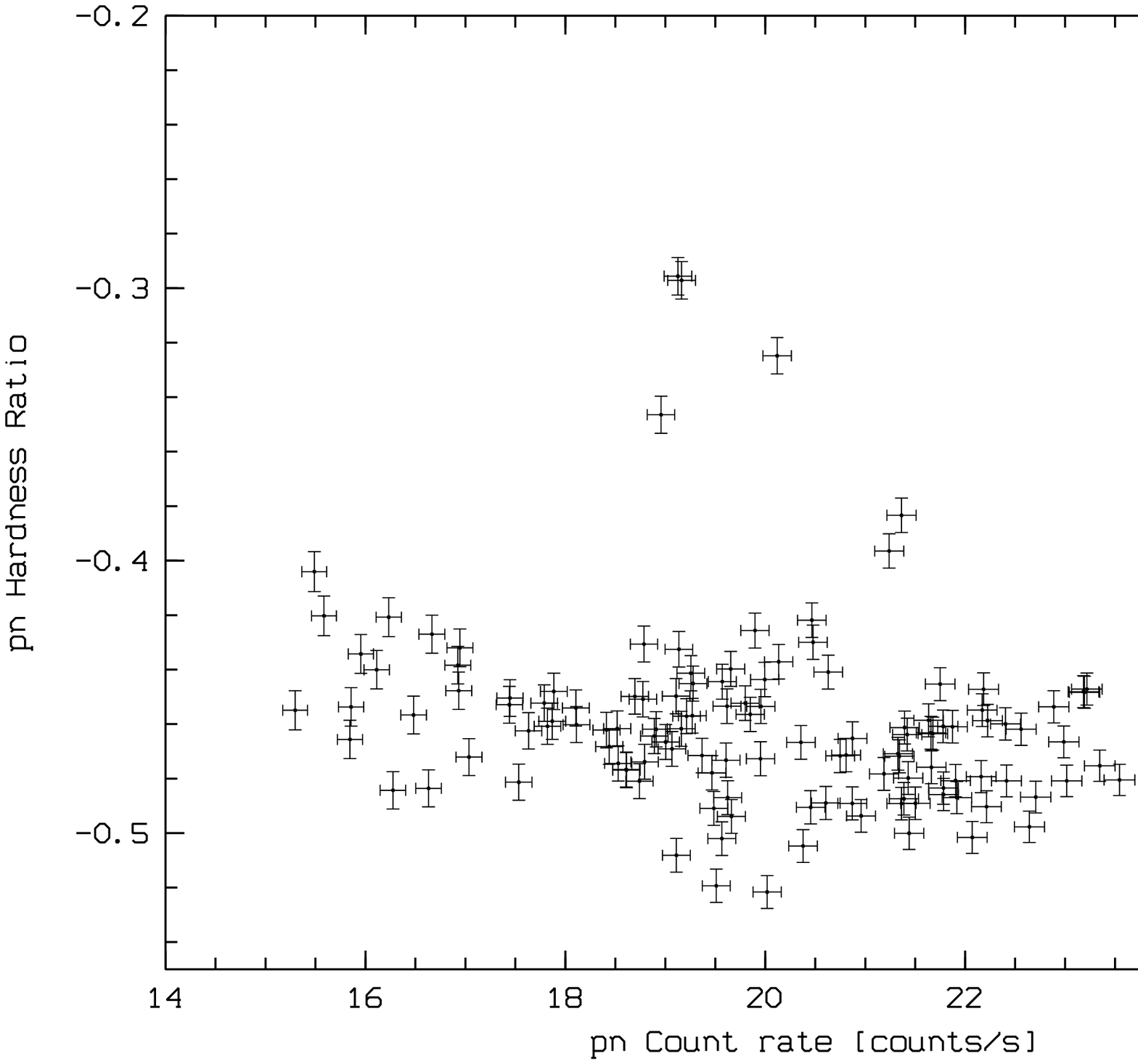}
\caption{Count rate vs. Hardness ratio of the 2009 (left) and 2006 (right)
 \xmm\ observations. \label{mkn335_xmm_cr_hr}
}
\end{figure*}

The \xmm\ 2009 pn light curve shown in Figure\,
\ref{mkn335_xmm_lc_2009} suggests
a dependence of the hardness ratio and therefore the shape of the X-
ray spectrum
on count rate. Figure\,\ref{mkn335_xmm_cr_hr} displays the count  
rate vs.
hardness ratios in the 2009 and 2006 \xmm\ observations in the left  
and right
panels, respectively. Clearly, there is a strong correlation between  
count rate
and hardness ratio in the low-state 2009 observation. A Spearman rank  
order test
results in a correlation coefficient $r_s=-0.73$ with a Student's T-
test value
$T_s=-15.3$ with a probability of P$<10^{-4}$ of a random result.  
However, the
high state data from 2006 give a completely different picture. Here we  
only see a
marginal trend that the source becomes softer with increasing count  
rate. A
Spearman rank order test results is $r_s=-0.20$, $T_s=-2.3$, and a  
probability
P=0.023.

The \swift\ long-term monitoring
data confirm that there is only a strong correlation between
the count rate and hardness ratio when the AGN is in the low state.
  This result is 
similar to the one found in the \xmm\ data. During the high state (like in the 2006  
\xmm\
data) we do not see a dependence of the X-ray spectral shape with X-
ray flux.

\subsection{X-ray spectral analysis}

\subsubsection{General remarks}
As shown in the previous subsection, Mkn 335 displays a strong  
dependence of the
shape of its X-ray spectrum on its X-ray flux:
the X-ray spectrum appears to be
harder when Mkn 335 is in a low state and softer when in a bright state.
Therefore performing and interpreting the spectral fits to the average data set will be  
limited.
We therefore split the data into 'bright' and faint'
states, defining 'bright' as phases when the pn count rate was 
$>$ 4   counts
s$^{-1}$ and 'faint' when the count rate was $<$ 3 counts s$^{-1}$,  
corresponding to
0.2-10 keV fluxes of 1.6$\times 10^{-14}$ and 1.2$\times 10^{-14}$ W m
$^{-2}$, respectively.
This results in four spectra, one 'bright' and 'faint' in each of the two  
orbits. 
However, most relevant are the spectra of the 'faint' state of the first orbit 
and the 'bright' state of the second orbit. The other two spectra do not have 
enough quality to allow detecting significant spectral changes. Therefore for
the remainder of the paper we focus on these two epochs of the 2009 observation.
The 2009 first orbit 'faint' state and the second orbit 'bright' state data are denoted 
(I) and (II) in Table\,\ref{xmm_spec_analysis}.
In
addition, we have also included the low-state spectrum from July 2007
\citep{grupe08a} to our analysis in order to  investigate, if and how the overall
spectral shape changed over those two years. This spectrum is marked as (III) in Table\,\ref{xmm_spec_analysis}.

\begin{figure*}
\epsscale{1.5}
\plottwo{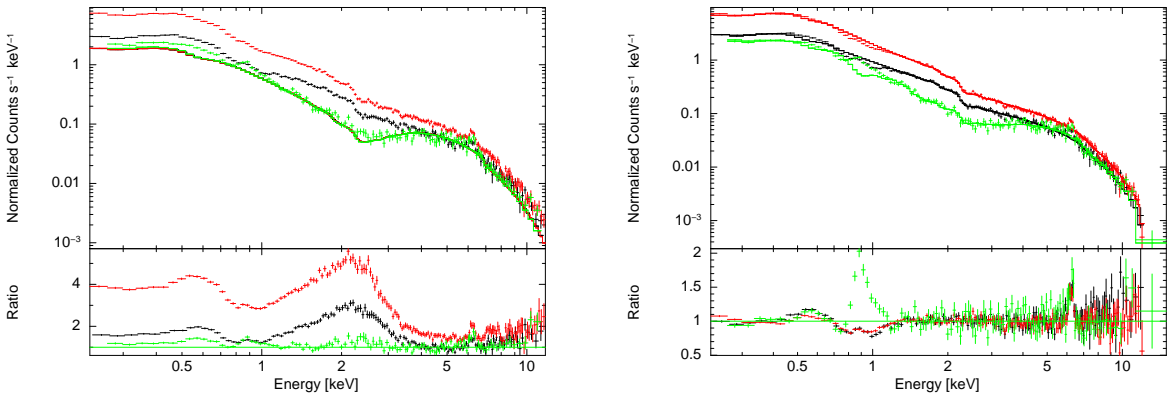}{f08b.ps}
\caption{\xmm\  2009 first orbit 'faint state', second orbit
'bright state' and the pn 2007 low state data of Mkn 335. The data are displayed in black, red and green,
respectively. In the left panel the spectra were modeled by a neutral partial covering absorber model using the
parameters as given in \citet{grupe08a}; $N_{\rm H,pc}=15.1\times 10^{22}$
cm$^{-2}$, $f_{\rm pc}$=0.94, \ax=1.78. In the right panel the spectra were fitted with an ionized partial covering absorber and an underlying black
body plus power law spectrum as listed in Table\,\ref{xmm_spec_analysis}.
 \label{mkn335_pn_2007_2009_plot}
}
\end{figure*}

In order to see how much the spectra have changed between  July 2007  
when Mkn 335 was in its low state \citep{grupe08a}
and  2009 June, as a first step, we fitted the new data with the 
2007 best-fit absorption model.  
The 2009 first orbit 'faint' and second  
orbit 'bright' mode were therefore described with   
  a power law model with Galactic and intrinsic
partial covering absorption, fixed to the parameters determined for the  
2007
low-state data
with $N_{\rm H, pc}=15.1\times 10^{22}$ cm$^{-2}$, $f_{\rm pc}$=0.94,  
and
\ax=1.78 \citep{grupe08a}. The absorber at z=0
was fixed to the Galactic value \citep[$3.96\times 10^{20}$
cm$^{-2}$;][]{dic90} which we used for all fits.
This fit is shown in the left panel of Figure\,
\ref{mkn335_pn_2007_2009_plot}
with the first orbit 'faint' spectrum displayed in black, the 'bright' second  
orbit spectrum
in red and the 2007 low state spectrum in green.
Clearly there is a strong deviation
of the 2009 data from the neutral partial covering absorber model used  
for the 2007
spectrum. This result suggests that the absorber and/or the intrinsic  
continuum spectrum must have
changed significantly. 

Therefore, in order to start exploring which model can best characterize the observed  
changes in the spectra, and which models can be safely ruled out, in a second step,
we  
applied some simple spectral models to the new data; and we continue to compare
with the previous data (note that those did not have simultaneous {\em deep} 
RGS observations).  
In these  
models we fix and tie
as many parameters as possible and then thaw them in order to study  
systematically the influence on the
spectral fits of each of these parameters.
There are strong residuals around 0.5, 0.9 and 6.4
keV which are likely due to well-localized 
  X-ray emission lines and absorption edges.
  In order to
constrain the  broadband continuum parameters,  
 we excluded the energy ranges 0.45-0.6, 0.7-1.1, and
5.5-6.7 keV at first from further analysis at this point. These energy bands
correspond to the OVII and OVIII emission lines and absorption edges, and the 
Fe K$\alpha$ emission line complex.  
This strategy will keep the number of free
parameters low and we can focus on the continuum properties first.

\subsubsection{Simple spectral models}

The results of 5 spectral fits
are summarized in Table\,\ref{xmm_spec_analysis}. We start 
with those simple models that have been routinely applied to essentially
all AGN observed so far: a single power law, a broken power law,
a power law with soft excess, and a power law with absorption. 
 First of all, we find that  a single
absorbed power law model does not result in an acceptable fit for any  
of these
spectra.  Although a broken power law model
does significantly improve the fits, it is not an acceptable
model for any of the spectra, either.
The same holds true for a power law plus black-body-type soft excess.   
Therefore, other models are required
to describe the data, and we continue with the next most obvious addition:
an ionized absorption component.  

\subsubsection{X-ray continuum fits with partial covering absorber  
models}

As a first step in characterizing the 2009 spectra,
we used a power law with  
neutral partial covering
absorber,  as we successfully applied it to the 2007 low state  
data. The absorption column density and the
covering fraction were left free to vary.
  However, this model
does not yield acceptable results for the 2009 data, neither when  
fitting the model to the
single spectra, nor when fitting it simultaneously to the 'faint state'  
of the first orbit and
the 'bright state' during the second orbit. 
 The next step was to fit the spectra  with an  
ionized partial
covering absorber \citep[$zxipcf$ in XSPEC as described by][]
{reeves08} and a
power law model. As shown in Table\,\ref{xmm_spec_analysis},
the underlying
intrinsic continuum spectrum can not be modeled by a single power law  
and requires an extra component, but addition of a black body component
improves the fit. 
  All spectra can be basically fitted
with an intrinsic spectrum with the same blackbody temperature 
and a hard X-ray spectral slope \ax=1.0.
When all parameters of the ionized partial
covering absorber are left free to vary, all 2009 spectra
show very similar covering
fractions and ionization  parameters. 
If we fix the
covering fraction to 51\% and the ionization parameter to
$log \xi$=1.91 [$10^{-5}$ W m, ergs s$^{-1}$ cm], then the
differences in the spectra are mostly due to changes in the absorption  
column
density of the ionized partial covering absorber. 

In a final step,  we  fitted  the 2009 'faint' and 'bright' spectra and
the 2007 low state spectrum
simultaneously, again with an ionized partial coverer.
As shown in Table\,\ref{xmm_spec_analysis}, the 2007 data can be fitted by the
ionized partial covering absorber model. However, they are fully consistent with
a neutral partial covering absorber model as well.

Given the possible degeneracy of the black-body component and the parameters
of the ionized absorber, further modeling and an  accurate parameter determination of
the absorber parameters, is not possible with the CCD-type spectra discussed here. Therefore, no
further modeling is presented here. 
%
In fact, our RGS analysis (in prep) suggests that a {\em multi}-component
absorber is preferred to fully characterize the ionized medium.  We have demonstrated that 
fitting the CCD spectrum with a single warm absorber is sufficient and more 
complex models are not warranted statistically, and would not yield meaningful results.

\subsubsection{Fits with Refection Models}
Although the continuum can be fitted by an ionized absorber model
quite well,
the previous spectral data of Mrk 335 could also be described in terms of the
blurred reflection model \citep[e.g.][]{ross05}.  
For completeness, here, we briefly show that such a model can also 
explain the new 2009 data; but we leave a study of the full parameter
space of possible models to a dedicated future study (Gallo et al., in
preparation).   


The initial model is the double reflector model used to interpret the
2007 X-ray weak state of Mrk 335 \citep{grupe08a}.  The model
considered the possibility of having the disc illuminated by two
different primary emitters; for example a compact emitter located
close to the black hole and a second, more extended corona
illuminating the disc at larger distances. In the 2007 low-state the
spectrum was described as being reflection dominated where the direct
emission from the power law component was significantly suppressed
relative to the reflection component.  An additional component
required in the 2007 low-state was emission from a distant ionized
emitter \citep{grupe08a, longinotti08}.  This was modeled using the
vmekal
model in XSPEC for a hot, diffuse gas.  There were no obvious
absorption features in the 2007 spectrum.

Here, the 2007 data and the 2009 data are fitted
together with the model described above. We find that the primary difference in the
continuum between 2007 and 2009 is the level of the power law emitter.
That is, the power law is more dominant in the 2009 date than in the
2007 low-state.  The ionized emitter remains constant in all three
spectra and is consistent with being emitted from large distances.
The apparent weakness of the emission spectrum in 2009 is attributed
to the increased fraction of the power law component in the X-ray
band.

Residuals remain in the fit at approximately 1.5 keV.  Considering contribution from
a warm absorber as is evident in the RGS analysis 
(Longinotti et al., in preparation) improves the fit.
In an upcoming work we are examining this model in much greater
detail by considering also the
variability of the source during each observation.

\subsubsection{Fe K$\alpha$ emission}

So far, the energy bins including the Fe K$\alpha$ emission line energy range
had been excluded from spectral fitting. The rest frame 6.4 keV Fe K$\alpha$
line is present in all spectra. The width of the line is about 
$\sigma$=140 eV.
In order to determine the flux and equivalent width of the Fe K$
\alpha$ line during each observation, we fitted
the spectra with a single power law plus redshifted Gaussian line in  
the 2-10 keV energy range.
We found that the line fluxes
determined from the 2009 'faint and 'bright' state and 2007 low state
spectra are (9.6\plm2.2)$\times 10^{-17}$ W m$^{-2}$,
(12.9\plm2.4)$\times 10^{-17}$ W m$^{-2}$, and (13.2\plm3.3)$\times  
10^{-17}$ W m$^{-2}$,
respectively.  
These fluxes suggest that the line has been constant regardless in which state
the AGN is.
The equivalent widths were 220\plm75, 200\plm60, and  
310\plm110 eV,
respectively. The values for the narrow Fe K$\alpha$ line are very similar  
when the reflection model is applied to the data.

\section{\label{discuss} Discussion and Conclusions}

We presented the results from our long-term monitoring campaign with  
\swift\ and the
short-term light-curve and X-ray spectral analysis of the highly  
variable NLS1 Mkn 335 using a dedicated, triggered 200 ks
observation with \xmm. Mkn 335 is one of the best examples of a  
typically bright AGN that goes through
states of low X-ray fluxes. Another example is the Seyfert 1 PG  
0844+349 for which we recently reported
on an \xmm\ observation during its deep low X-ray flux state  
\citep{gallo10}. After Mkn 335 was discovered in an extremely
low X-ray flux state in May 2007 we discovered in the 20 ks \xmm\  
observation from July 2007 that it showed strong soft
X-ray emission lines \citep{grupe08a}. This observation however was  
too short to put constraints on the ionized gas properties
and to model the continuum shape of the low-state in detail.
Therefore, we triggered the deep \xmm\ observation discussed here,
which also led to the first detection of narrow absorption
lines in Mkn 335 with RGS (Longinotti et al., in preparation).

\subsection{Continuum spectrum}

As we saw previously for the 2007 low state \xmm\ observations of Mkn 335,  
partial covering absorber and reflection models
yield a similar quality of the spectral fits.
The continuum
spectrum is highly variable and complex. When Mkn 335 was observed by  
\xmm\ in June 2009 it varied very fast on timescales of just
hours with the spectrum becoming softer with increasing X-ray flux.  
The 2009 data require that the absorber has to be
ionized. 
Since there were no signs of intrinsic absorption features in
the 2000 and 2006 \xmm\ and {\it Suzaku} data, 
the presence of the absorber is not permanent, but transient.

 How is a partial coverer model consistent with the long-term variability 
behavior of this NLS1 seen by {\it Swift} ?    
 As shown in Sections \ref{var_long} and \ref{var_short}
the X-ray spectra become softer with increasing X-ray flux. 
At first glance, an  
increase in flux could mean intrinsically an increase of the
accretion rate and therefore luminosity and
$L/L_{\rm Edd}$ which would result in a steeper X-ray spectrum
\citep{grupe04b, grupe10,
shemmer08}. 
However, as we have seen from the shape of the X-ray  
spectrum, this  simple picture can not explain the X-ray spectrum which
is much more complicated than a simple power law model. A variable  
partial covering absorber, however, can explain the X-ray light
curves seen on long as well and short time scales: When the absorber  
becomes stronger and the observed X-ray flux lower, the spectrum
becomes harder. On the other hand, when the absorber becomes more  
transparent or even disappears the spectrum becomes soft.

\subsection{X-ray variability} 
As we have shown in Figures\,\ref{mkn335_lc}, \ref{mkn335_xmm_lc_2009}, and
\ref{mkn335_xmm_lc_2006} Mkn 335 is highly variable in X-rays on long and short time
scales. Our long-term \swift\ monitoring has shown that Mkn 335 varies in X-rays by
factors of about 40 even within months. Beside this strong flux variability we also 
observe a strong spectral variability when the AGN is in a low state. Above a certain
threshold we only see X-ray flux variability with no significant changes in the hardness
ratio. This behavior appears on short as well as on long time scales (see
Figures\,\ref{mkn335_xrt_cr_hr} and \ref{mkn335_xmm_cr_hr}). Such a behavior has also been
reported by \citet{turner08, turner11} for NGC 3516. If we assume the partial covering
absorber picture, then the spectral
changes in the hardness ratio at lower X-ray fluxes can be explained
in terms of changes in the absorber
properties, such as the column density, covering fraction and ionization parameter.


Indeed, we find that the flux in the Fe K$\alpha$ line is constant, regardless 
of the state of the AGN. 
This constant line flux indicates an underlying 
continuum component that is not variable.

\subsection{UV variability}
The UV light curve of Mkn 335 is quite remarkable. While NLS1s typically vary by
only 0.3 mag in the UV \citep{grupe10}, Mkn 335 shows variability by about 1 mag
over a time scale of a few months as seen between 2010 September and December and 2011 June and September. 
(The strongest UV variability we have seen from the
\swift\ observations listed in \citet{grupe10}, 
was from Seyfert 1.5 galaxies). In Mkn 335, we
did not find any clear correlation between the X-ray and UV flux changes during
the time period 2007 May to 2011 July. 
This result may suggest that the changes in the UV flux are not 
directly linked  to changes in
the X-ray flux. 
 However, as shown in the \xmm\ short-term light curve 
Figure\,\ref{mkn335_xmm_lc_2009} the brightening in the OM W2 light curve at the
end could be seen as a response to the 'flare' at 225 ks in X-rays. Currently we
do not have the temporal resolution to exclude that there is not a direct
connection between the UV and X-rays. Our investigation of possible lags between
the UV and X-ray emission is certainly affected by the under-sampling in the
\swift\ light curves. As we have seen from the \xmm\ 2009 light curve, Mkn 355
is highly variable even on time scales of hours. Our current \swift\ light curve
systematically under-samples these time scales.

\subsection{Mkn 335 and AGN in deep minimum X-ray flux states}
Mkn 335 is one of the best examples of an AGN that used to be typically in an X-ray
bright flux state but then suddenly becomes dramatically fainter. 
The most extreme of these cases is the Narrow-Line Seyfert 1 galaxy  (NLS1)
WPVS 007
\citep{gru95} which dropped by a factor of more than 400 between its
ROSAT All-Sky Survey \citep[RASS; ][]{voges99} and ROSAT pointed  
observations
about three years later. FUSE UV spectroscopy and \swift\ X-ray  
observations
revealed the presence of strong UV absorption line troughs and a partial
covering absorber in X-rays \citep{leighly09, grupe08b}. In the case of WPVS 007 our
interpretation is that this is a low-luminosity, low-redshift analogon of a Broad
Absorption Line Quasar (BAL QSO)
as suggested by \citet{leighly09}. Because the black hole mass in
WPVS 007 is only a few 10$^6$ \msun, the timescales in this AGN are hundreds of times
shorter than in a typical BAL QSO. Mkn 335 could be just another example of such a BAL QSO
analogons. As pointed out by \citep{brandt00} and \citet{boroson02}, BAL QSOs and NLS1s are
both AGN with high $L/L_{\rm Edd}$
 Eddington ratios, but with significantly different central black
hole masses. WPVS 007 as well as Mkn 335 may be the link between these two AGN classes.

\subsection{Conclusions}

We presented the results from a more than four year long monitoring  
campaign with \swift\  of the Narrow Line Seyfert 1 galaxy Mkn  
335, one of the longest with simultaneous X-ray
and UV measurements so far obtained for an AGN. We also presented  a
200 ks observation with \xmm\, triggered at low flux state.
We found that

\begin{itemize}
\item Mkn 335 continues to be highly variable in X-rays on long and  
short time scales. The total amplitude of variability in count rate
in the \swift\ XRT data 
(peak to dip between 2007 and 2011) is a factor of 24. The lowest count rate was
seen on 2011 August 28 with 0.042 counts s$^{-1}$ and the highest count rate on
2007 October 18 with 1.277 counts s$^{-1}$.
The fastest
doubling timescale we saw from the \xmm\ observation in 2009 was about 2
hours during the first orbit making it one of the most rapidly varying AGN in
X-rays.

\item The X-ray and UV variability is not strongly correlated.
However, during X-ray bright states, the faintest UV states do not
occur, while during X-ray low-states the amplitude of UV variability
is highest. 

\item With a variability of about 1 mag in the UV within a few months as seen
between 2011 June and September,
Mkn 335 is one of the most variable NLS1 known in the UV. 

\item Its X-ray hardness ratio shows distinct variability patterns
in high- and low-state. During the low-states, there is a clear correlation
of hardness ratio with count rate. However, this pattern disappears in high-state; 
hardness ratio is independent of count rate, and occasionally shows some 
abrupt changes on short time scales.   

\item Formally, both
ionized absorbers and blurred reflectors do provide
successful spectral fits to the XMM low-state data.   

\end{itemize}

While the presence of ionized absorption is confirmed by the RGS data (in prep.),
the number and the properties of all spectral components present at any given
time will be addressed by in-depth
follow-up modeling. 

Mrk 335 continues to be one of the few AGN which are still bright enough in their low-states 
for spectral analysis, and therefore hold the best hopes of understanding 
AGN spectral components,
spectral complexity, and mechanisms of variability. 
We continue to monitor Mrk 335, in order to identify
 pronounced high- and low-states.


%


\acknowledgments
We thank the \swift\ PI Neil Gehrels for approving our various
ToO requests to monitor Mkn 335 with \swift, and the \xmm\ Science  
Operations Team
for their fast turn around when we requested our 200 ks triggered observation.
Many thanks also to the anonymous referee for useful comments and suggestions.
We acknowledge the use of public data from the Swift data archives.
This research has made use of the NASA/IPAC Extragalactic
Database (NED) which is operated by the Jet Propulsion Laboratory,
Caltech, under contract with the National Aeronautics and Space
Administration. This research has made use of the
  XRT Data Analysis Software (XRTDAS) developed under the responsibility
  of the ASI Science Data Center (ASDC), Italy.
SK acknowledges
the hospitality of the Aspen Center for Physics, and thanks the participants
for discussions on AGN X-ray spectral models. 
DX and SK acknowledge support from the Chinese National 
Science Foundation (NSFC) under grant NSFC 10873017. DX also
acknowledges support from program 973 (2009CB824800).
  Swift is supported at PSU by NASA contract NAS5-00136.
This research was supported by NASA contracts NNX08AT25G,  NNX09AP50G, and
NNX09AN12G (D.G.).

\clearpage


\begin{deluxetable}{cccccrrrrrrr}
\tabletypesize{\tiny}
\tablecaption{\swift~ Observation log of Mkn 335
\label{swift_log}}
\tablewidth{0pt}
\tablehead{
\colhead{ObsID} &
\colhead{Segment} &
 \colhead{T-start\tablenotemark{1}} & 
\colhead{T-stop\tablenotemark{1}} 
& \colhead{MJD\tablenotemark{2}} 
& \colhead{$\rm T_{\rm XRT}$\tablenotemark{3}} 
& \colhead{$\rm T_{\rm V}$\tablenotemark{3}} 
& \colhead{$\rm T_{\rm B}$\tablenotemark{3}} 
& \colhead{$\rm T_{\rm U}$\tablenotemark{3}} 
& \colhead{$\rm T_{\rm UVW1}$\tablenotemark{3}} 
& \colhead{$\rm T_{\rm UVM2}$\tablenotemark{3}} 
& \colhead{$\rm T_{\rm UVW2}$\tablenotemark{3}} 
} 
\startdata
35755 & 001 & 2007-05-17 00:32 & 2007-05-17 05:37 & 54237.128 &   4859 &  401 &  401 &  401 &  800 & 1082 & 1603 \\ 
      & 002 & 2007-05-25 00:01 & 2007-05-25 19:23 & 54245.406 &   8084 &  658 &  658 &  658 & 1323 & 1677 & 2648 \\ 
      & 003 & 2007-06-28 00:01 & 2007-06-28 14:37 & 54279.073 &   2932 &  273 &  273 &  273 &  545 &  677 & 1092 \\ 
      & 004 & 2007-06-30 00:13 & 2007-06-30 14:52 & 54281.314 &   2837 &  230 &  230 &  230 &  461 &  615 &  921 \\ 
      & 005 & 2007-07-02 14:47 & 2007-07-02 21:18 & 54283.751 &   2979 &  245 &  245 &  245 &  490 &  614 &  983 \\ 
      & 008 & 2007-09-19 00:05 & 2007-09-19 11:24 & 54362.219 &   1389 & \nodata & \nodata & \nodata & \nodata & \nodata & \nodata \\
      & 009 & 2007-09-25 00:55 & 2007-09-25 05:51 & 54368.141 &   2050 & \nodata & \nodata & \nodata & \nodata & \nodata & \nodata \\
      & 010 & 2007-10-04 00:01 & 2007-10-04 03:13 & 54377.068 &   1401 & \nodata & \nodata & \nodata & \nodata & \nodata & \nodata \\
\enddata

\tablenotetext{1}{Start and End times are given in UT}
\tablenotetext{2}{The MJD marks the middle of the observation}
\tablenotetext{3}{Observing time given in s}
\end{deluxetable}

\begin{deluxetable}{lccrcc}
\tabletypesize{\tiny}
\tablecaption{\xmm~ Observation log of  Mkn 335
\label{xmm_log}}
\tablewidth{0pt}
\tablehead{
\colhead{Filter/Detector} &
 \colhead{T-start\tablenotemark{1}} & 
 \colhead{T-end\tablenotemark{1}} & 
\colhead{T-exp\tablenotemark{2}} 
& \colhead{Mag\tablenotemark{3}} 
& \colhead{2007 Mag\tablenotemark{4}} 
} 
\startdata
pn    & 2009-06-11 07:39 & 2009-06-12 19:54 & 114784 & \\
MOS-1 & 2009-06-11 07:16 & 2009-06-12 19:57 & 131920 & \\
MOS-2 & 2009-06-11 07:17 & 2009-06-12 19:57 & 131972 & \\
pn    & 2009-06-13 07:31 & 2009-06-14 05:56 &  80344 & \\
MOS-1 & 2009-06-13 07:08 & 2009-06-14 06:00 &  82257 & \\
MOS-2 & 2009-06-13 07:08 & 2009-06-14 06:00 &  82249 & \\
OM V  & 2009-06-11 07:25 & 2009-06-11 08:38 & 4400 & 14.47\plm0.01 \\
OM U  & 2009-06-11 08:43 & 2009-06-11 09:57 & 4400 & 13.65\plm0.01 & 13.36\plm0.01 \\
\enddata

\tablenotetext{1}{Start and end times are given in UT}
\tablenotetext{2}{Exposure times given in s}
\tablenotetext{3}{Observed magnitudes not corrected for Galactic reddening at the
position of Mkn 335}
\tablenotetext{4}{The 2007 OM data are listed here as a comparison to the 2009 observations. For the exact observing time we refer to \citet{grupe08a}.
}
\end{deluxetable}

\begin{deluxetable}{ccccccccccc}
\tabletypesize{\tiny}
\tablecaption{\swift~XRT count rates and hardness ratios and UVOT Magnitudes 
of Mkn 335
\label{swift_results}}
\tablewidth{0pt}
\tablehead{
\colhead{ObsID}
& \colhead{Segment}
& \colhead{MJD}
& \colhead{XRT CR} 
& \colhead{XRT HR} 
& \colhead{V} 
& \colhead{B} 
& \colhead{U}  
& \colhead{UVW1} 
& \colhead{UVM2} 
& \colhead{UVW2}  
} 
\startdata
35755 & 001 & 54237.128 &  0.053\plm0.004 & $+$0.07\plm0.03 & 14.22\plm0.01 & 14.49\plm0.01 & 13.29\plm0.01 & 13.20\plm0.01 & 13.12\plm0.01 & 13.19\plm0.01 \\ 
      & 002 & 54245.406 &  0.144\plm0.004 & $+$0.06\plm0.03 & 14.19\plm0.01 & 14.42\plm0.01 & 13.20\plm0.01 & 13.08\plm0.01 & 12.95\plm0.01 & 13.01\plm0.01 \\ 
      & 003 & 54279.073 &  0.076\plm0.012 & $+$0.10\plm0.07 & 14.19\plm0.01 & 14.44\plm0.01 & 13.23\plm0.01 & 13.09\plm0.01 & 12.98\plm0.01 & 13.04\plm0.01 \\ 
      & 004 & 54281.314 &  0.082\plm0.006 & $+$0.04\plm0.07 & 14.15\plm0.01 & 14.41\plm0.01 & 13.22\plm0.01 & 13.10\plm0.01 & 12.97\plm0.01 & 13.04\plm0.01 \\ 
      & 005 & 54283.751 &  0.109\plm0.007 & $-$0.15\plm0.06 & 14.18\plm0.01 & 14.42\plm0.01 & 13.23\plm0.01 & 13.05\plm0.01 & 12.93\plm0.01 & 12.98\plm0.01 \\ 
      & 008 & 54362.219 &  0.420\plm0.020 & $-$0.15\plm0.05 &\nodata & \nodata & \nodata & \nodata & \nodata & \nodata \\
      & 009 & 54368.141 &  0.634\plm0.021 & $-$0.07\plm0.03 &\nodata & \nodata & \nodata & \nodata & \nodata & \nodata \\
      & 010 & 54377.068 &  0.154\plm0.011 & $+$0.25\plm0.07 &\nodata & \nodata & \nodata & \nodata & \nodata & \nodata \\
\enddata
\tablenotetext{1}{Magnitude corrected for reddening with $E_{\rm B-V}$=0.037 given by
\citet{sfd98}. The errors given in this table are statistical errors}
\end{deluxetable}

\begin{deluxetable}{llcccccccc}
\tabletypesize{\tiny}
\tablecaption{Results of the spectral fits to the 'faint' and 'bright' state
\xmm\ pn data 
 of Mkn 335 during the first and second orbit.
\label{xmm_spec_analysis}}
\tablewidth{0pt}
\tablehead{
 \colhead{Spectrum\tablenotemark{1}} &
 \colhead{Model\tablenotemark{2}} &
 \colhead{$\alpha_{\rm X,soft}$} & 
 \colhead{$E_{\rm break}$} & 
 \colhead{$\alpha_{\rm X,hard}$} 
& \colhead{$T_{\rm BB}$\tablenotemark{3}}
& \colhead{$N_{\rm H,pc}$\tablenotemark{4}}
& \colhead{$f_{\rm pc}$}
& \colhead{log($\xi$)\tablenotemark{5}}
& \colhead{$\chi^2/\nu$}
} 
\startdata
(I) & powl & 1.36\plm0.01 & --- & --- &  --- & --- & --- & --- & 4071/117 \\
& bknpo & 1.74\plm0.02 & 1.27\plm0.03 & 0.52\plm0.02 &  --- & --- & --- & ---  & 198/115 \\
& bb + po & --- & --- & 0.52\plm0.02 & 99\plm2 & --- & --- & --- & 209/115 \\
& zpcfabs * po & 1.65\plm0.01 & --- & --- & --- & 5.1\plm0.2 & 0.78\plm0.01 & --- & 516/115 \\
& zpcfabs * (bb + po) & --- & --- & 0.77$^{+0.06}_{-0.12}$ & 97\plm2 & 3.7\plm1.0 & 0.33$^{+0.05}_{-0.13}$ & --- & 191/113 \\
& zxipcf * po & 1.18\plm0.01 & --- & --- & --- & 5.75\plm0.30 & 1.00\tablenotemark{6} & 2.24\plm0.02 & 255/114 \\
& zxipcf * (bb+po) & --- & --- & 0.92$^{+0.05}_{-0.10}$ & 110$^{+3}_{-5}$ & 10.5\plm1.5 & 0.51$^{+0.05}_{-0.09}$ &
1.91$^{+0.05}_{-0.08}$ & 154/112 \\
& & \\
\enddata
\tablenotetext{1}{Spectra: (I) 2009, Orbit 1, faint; (II)  2009, Orbit 2,
bright; (III) 2007}
\tablenotetext{2}{Spectra models: powl: single power law; bknpo: broken power law; bb: black body; 
zpcfabs: redshifted neutral partial covering absorber,
 zxipcf: ionized redshifted partial covering absorber; 
Note that all models are fitted with the absorption model wabs with the column density fixed to
the Galactic value of 3.96$\times 10^{20}$ cm$^{-2}$ taken from \citet{dic90}. For all models we excluded the energy ranges 0.45-0.6,
0.7-1.1, and 5.5-6.7 keV.}
\tablenotetext{3}{The black body temperature $T_{\rm BB}$  is given in units of eV.}
\tablenotetext{4}{The column density of the ionized partial covering absorber is given in unites of $10^{22}$ cm$^{-2}$.}
\tablenotetext{5}{The ionization parameter $\xi$ is given in units $10^{-5}$ W m, or ergs s$^{-1}$ cm. }
\tablenotetext{6}{The covering fraction of the ionized absorber pegged.}
\tablenotetext{7}{Covering fraction and ionization parameter fixed to the values of the low state. When left as free parameters log $\xi$ results in the
same value, but $f_{\rm pc}$ is not constrained.}
\end{deluxetable}


\begin{thebibliography}{}
\bibitem[Albert (2009)]{albert09} Albert, J., 2009, ``Bayesian Computation with
R'', Springer
\bibitem[Arevalo et al(2008)]{arevalo08} Arevalo, P., et al., 2008,  
    \mnras, 388, 211 
\bibitem[Arnaud (1996)]{arnaud96} Arnaud, K.~A., 1996, ASP
Conf.~Ser.~101: Astronomical Data Analysis Software and Systems V,  
101, 17
\bibitem[Ballantyne et al.(2001)]{ballantyne01} Ballantyne, D.R.,  
Iwasawa, K., \& Fabian, A.C., 2001, \mnras, 323, 506
\bibitem[Ballo et al.(2008)]{ballo08} Ballo, L., et al., 2008, \aap,  
483, 137
\bibitem[Boroson (2002)]{boroson02} Boroson, T.A., 2002, \apj, 565, 78
\bibitem[Brandt \& Gallagher (2000)]{brandt00} Brandt, W.N., \& Gallagher, S.C., 2000, New
Astronomy Review, 44, 461
\bibitem[Breeveld et al.(2010)]{breeveld10} Breeveld, A.A., et al.,  
2010, \mnras, 406, 1687
\bibitem[Burrows et al. (2005)]{burrows04} Burrows, D., et al., 2005,  
Space Science Reviews, 120, 165
\bibitem[Cardelli et al.(1989)]{cardelli89} Cardelli, J.A., Clayton,  
G.C., Mathis, J.S., 1989, \apj, 345, 245
\bibitem[Chevalier et a.(2006)]{chevalier06} Chevalier, L., Collins,  
S., Dumont,
A.-M., Czerny, B., Mouchet, M., Gon\c{c}alves, A.C., \& Goosmann, R.,  
2006,
\aap, 449, 493
\bibitem[den Herder et al.(2001)]{denherder01} den Herder, J.W., et  
al., 2001,
\aap, 365, L17
\bibitem[Dickey \& Lockman (1990)]{dic90} Dickey, J.M., \& Lockman,  
F.J., 1990,
\araa, 28, 215
\bibitem[Fabian et al.(2002)]{fabian02} Fabian, A.C., et al., 2002,  
\mnras, 331,
L35
\bibitem[Fabian et al.(2004)]{fabian04} Fabian, A.C., Miniutti, G.,  
Gallo, L.C.,
Boller, T., Tanaka, Y., Vaughan, S., \& Ross, R.R., 2004. \mnras, 353,  
1071
\bibitem[Fabian et al.(2009)]{fabian09} Fabian, A.C., et al., Nature,  
459, 540
\bibitem[Feroz et al.(2009)]{feroz09} Feroz, F., Hobson, M.P., \&  
Bridges, M.,
2009, \mnras, 398, 1601
\bibitem[Gallo (2006)]{gallo06} Gallo, L.C., 2006, \mnras, 368, 479
\bibitem[Gallo et al.(2011)]{gallo10} Gallo, L.C., Grupe, D.,  
Schartel, N., Komossa,
S., Miniutti, G., Fabian, A.C., \& Santos-Lleo, M., 2011,
\mnras, 412, 161
\bibitem[Gehrels et al. (2004)]{gehrels04} Gehrels, N., et al., 2004,  
ApJ, 611,
1005
\bibitem[George et al.(2000)]{george00} George, I.M., Turner, T.J.,  
Yaqoob, T., Netzer, H., Laor, A., Mushotzky, R.F., Nandra, K.,
\& Takahashi, T., 2000, \apj, 531, 52
\bibitem[Godet et al. (2009)]{godet09} Godet, O., et al., \aap, 494, 775
\bibitem[Gondoin et al. (2002)]{gondoin02} Gondoin, P., Orr, A., Lumb,  
D., \&
Santos-Lleo, M., 2002, \aap, 388,74
\bibitem[Gregory (2005)]{gregory05} Gregory, P.C., 2005, \apj, 631, 1198
\bibitem[Grier et al.(2012)]{grier11} Grier C.J., et al., 2012, \apj, 744, L4
\bibitem[Grupe et al. (1995)]{gru95} Grupe, D., Beuermann, K.,  
Mannheim, K.,
Thomas, H.-C., de Martino, D., \& Fink, H.H., 1995, \aap, 300, L21
\bibitem[Grupe (2004)]{grupe04b} Grupe, D., \aj, 127, 1799
\bibitem[Grupe et al.(2001)]{grupe01} Grupe, D., Thomas, H.-C., \&  
Beuermann, K.,
2001, \aap, 367, 470
\bibitem[Grupe et al. (2004a)]{grupe04} Grupe, D., Wills, B.J.,  
Leighly, K.M., \&
Meusinger, H., 2004a, \aj, 127, 156
Komossa, S., 2004b, \aj, 127, 3161
\bibitem[Grupe et al.(2007a)]{grupe07a} Grupe, D., Schady, P.,  
Leighly, K.M.,
  Komossa, S., O'Brien, P.T., \& Nousek, J.A., 2007, \aj, 133, 1988
\bibitem[Grupe et al.(2007b)]{grupe07b} Grupe, D., Komossa, S., \&  
Gallo, L.,
2007b, \apj, 668, L111
\bibitem[Grupe et al.(2008a)]{grupe08a} Grupe, D., Komossa, S., Gallo,  
L.C.,
Fabian, A., Larrson, J., Pradhan, A.K., Xu, D., \& Miniutti, G.,  
2008a, \apj,
681, 982
\bibitem[Grupe et al.(2008b)]{grupe08b} Grupe, D., Leighly, K.M., \&  
Komossa,
S., 2008b, \aj, 136, 2343
\bibitem[Grupe et al.(2010)]{grupe10} Grupe, D., Komossa, S., Leighly,  
K.M., \&
Page, K.L., 2010, \apjs, 187,64
\bibitem[Halpern(1982)]{halpern82} Halpern, J.P., 1982, PhD thesis,  
Harvard University
\bibitem[Hill et al. (2004)]{hill04} Hill, J.E., et al., 2004, SPIE,  
5165, 217
\bibitem[Jansen et al.(2001)]{jansen01} Jansen, F., et al., 2001,  
\aap, 365, L1
\bibitem[Larsson et al.(2008)]{larsson08} Larsson, J., Miniutti, G.,  
Fabian, A.C., Miller,
J.M., Reynolds, C.S., \& Ponti, G., 2008, \mnras, 384, 1316
\bibitem[Leighly (1999)]{lei99b} Leighly, K.M., 1999, \apjs, 125, 317
\bibitem[Leighly et al.(2007)]{leighly07} Leighly, K.M., Halpern,  
J.P., Jenkins,
E.B., Grupe, D., Choi, J., \& Prescott, K.B.,   2007, \apj, 663, 103
\bibitem[Leighly et al.(2009)]{leighly09} Leighly, K.M., Hamann, F.,  
Casebeer,
D.A., \& Grupe, D., 2009, \apj, 701, 176
\bibitem[Longinotti et al.(2007a)]{longinotti07a} Longinotti, A.L.,  
Sim, S.A.,
Nandra, K., \& Cappi, M., 2007a, \mnras, 374, 237
\bibitem[Longinotti et al. (2007)]{longinotti07b} Longinotti, A.L.,  
Sim, S.A.,
Nandra, K., Cappi, M., \& O'Neill, P., 2007b, ASP Conf. Series, Vol. 373, 
proceedings of the conference held 16-21 October, 2006 at Xi'an Jioatong 
University, Xi'an, China. Edited by Luis C. Ho and Jian-Min Wang, p.341 
\bibitem[Longinotti et al. (2008)]{longinotti08} Longinotti, A.L.,  
Nucita, A., Santos-Lleo, M., \&
Guainazzi, M., 2008, \aap, 484, 311
\bibitem[Longinotti et al. (2009)]{longinotti09} Longinotti, A.L., Bianchi, S., Ballo, L.,
de la Calle, I., \& Guainazzi, M., 2009, \mnras, 394, L1
\bibitem[Mason et al. (2001)]{mason01} Mason, K.O., et al., 2001,  
\aap, 365, L36
\bibitem[Merloni et al.(2006)]{merloni06} Merloni, A., Malzac, J.,  
Fabian, A.C., \& Ross, R.R., 2006, \mnras, 370, 1699
\bibitem[Miller et al. (2007)]{miller07} Miller, L., Turner, T.J., Reeves, J.N., George,
I.M., Kraemer, S.B., \& Wingert, B., 2007, \aap, 463, 131
\bibitem[Miller et al.(2008)]{miller08} Miller L., Turner, T.J., \&  
Reeves, J.N., 2008, \aap, 483, 437
\bibitem[Miller et al.(2009)]{miller09} Miller L., Turner, T.J., \&  
Reeves,
J.N., 2009, \mnras, 399, L96
\bibitem[Miniutti et al.(2009)]{miniutti09} Miniutti, G., Fabian,  
A.C., Brandt,
W.N., Gallo, L.C., \& Boller, T., 2009, \mnras, 396, L85
\bibitem[Nahar et al.(2011)]{nahar11} Nahar, S.N., et al., 2011, Phys Rev A, 83,
053417
\bibitem[Nandra \& Pounds(1994)]{nandra94} Nandra, K., \& Pounds,  
K.A., 1994, \mnras, 268, 405
\bibitem[O'Neill et al.(2007)]{oneill07} O'Neill, P.M., Nandra, K.,  
Cappi, M.,
Longinotti, A.L., \& Sim, S.A., 2007, \mnras, 381, L94
\bibitem[Peterson et al.(2004)]{peterson04} Peterson, B.M., et al., 2004, \apj,
613, 682
\bibitem[Poole et al. (2008)]{poole08} Poole, T.S., et al., 2008,  
\mnras, 383, 627
\bibitem[Pounds et al.(1987)]{pounds87} Pounds, K.A., Stanger, V.J.,  
Turner, T.J., King, A.R., \& Czerny, B., 1987, \mnras, 224, 443
\bibitem[Reeves et al. (2008)]{reeves08} Reeves, J., et al., 2008,  
\mnras, 385,
L108
\bibitem[Risaliti et al.(2009)]{risaliti09} Risaliti, G., et al.,  
2009, \mnras,
393, L1
\bibitem[Roming et al. (2005)]{roming04} Roming, P.W.A., et al., 2005,  
Space Science Reviews, 120, 95
\bibitem[Roming et al.(2009)]{roming09} Roming, P.W.A., et al., 2009,  
\apj, 690, 163
\bibitem[Ross \& Fabian (2005)]{ross05} Ross, R.R., \& Fabian, A.C.,  
2005, \mnras, 358, 211
\bibitem[Schartel et al.(2007)]{schartel07} Schartel, N., et al.,  
2007, \aap,
474, 431
\bibitem[Schartel et al.(2010)]{schartel10} Schartel, N., Rodr\'iguez-
PAscual, P.M., Santos-Lle\'o, M.,
Jim\'enez-Bail\'on, E., Ballo, L., Piconcelli, E., 2010, \aap,  
512, 75
\bibitem[Shemmer et al.(2008)]{shemmer08} Shemmer, O., Brandt, W.N.,  
Netzer, H., Maiolino, R., \& Kaspi, S., 2008, \apj, 682, 81
\bibitem[Schlegel et al.(1998)]{sfd98} Schlegel, D.~J., Finkbeiner,  
D.~P.,
\& Davis, M.\ 1998, \apj, 500, 525
\bibitem[Str\"uder et al.(2001)]{strueder01} Str\"uder, L., et al.,  
2001, \aap, 365, L18
\bibitem[Tanaka et al. (1995)]{tanaka95} Tanaka, Y., et al., 1995,  
Nature, 375,
659
\bibitem[Tananbaum et al.(1978)]{tananbaum78} Tananbaum, H., Peters,  
G., Forman, W., Giacconi, R., Jones, C., \&
Avni, Y., 1978, \apj, 223, 74
\bibitem[Turner et al.(1993)]{turner93} Turner, T.J., et al., 1993,  
\apj, 407, 556
\bibitem[Turner et al.(2001)]{turner01} Turner, M.J., et al., 2001,  
\aap, 365,
L27
\bibitem[Turner et al. (2007)]{turner07} Turner, T.J., Miller, L., Reeves, J.N., \&
Kraemer, S.B., 2007, \aap, 475, 121
\bibitem[Turner et al. (2008)]{turner08} Turner, T.J., Reeves, J.N., Kraemer, S.B., \&
Miller, L., 2008, \aap, 483, 161
\bibitem[Turner et al. (2011)]{turner11} Turner, T.J., Miller, L., Kraemer, S.B., \&
Reeves, J.N., 2011, \apj, 733, 48
\bibitem[Turner \& Miller (2009)]{turner09} Turner, T.J., \& Miller,  
L., 2009, A\&AR, 17, 47
\bibitem[Vignali et al.(2008)]{vignali08} Vignali, C., et al., 2008,  
\mnras,
388, 761
\bibitem[Voges et al. (1999)]{voges99} Voges, W., Aschenbach, B.,  
Boller, T., et
al., 1999, \aap, 349, 389
\end{thebibliography}
\end{document}